**Text2Struct: A Machine Learning Pipeline for Mining Structured Data from Text**

Chaochao Zhou[1,2,*], Bo Yang[3]

[1]Department of Radiology, Northwestern University Feinberg School of Medicine;
[2]Department of Computer Science, University of Illinois Urbana-Champaign;
[3]LinkedIn

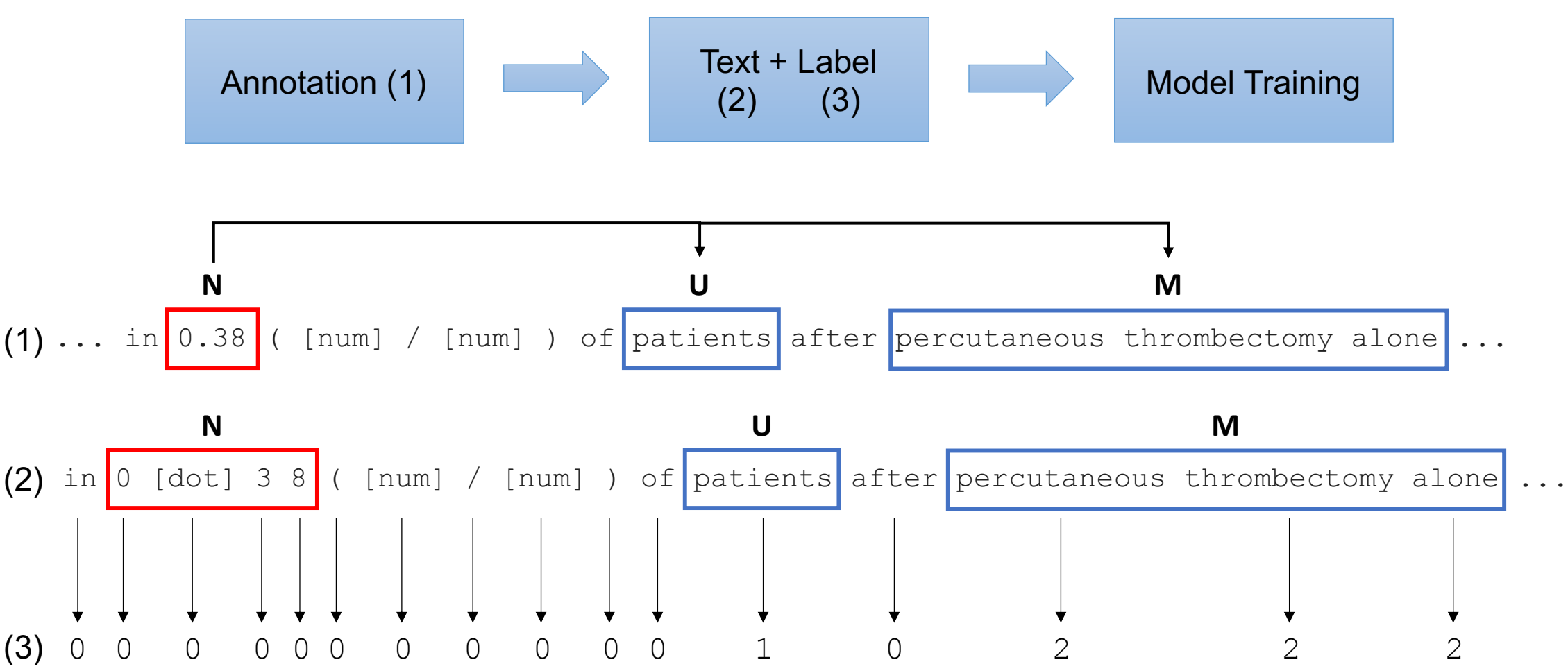

*An overview of the pipeline that processes a simple example from the annotation to the text-label pair used for model training. (1) Annotation of the unit (U) and metric (M) associated with a single numeral (N), while other numerals have been masked as "[num]". (2) The text in which the numeral is represented by unigram characters. (3) The resulting labels corresponding to each word (0 = none; 1 = unit; 2 = metric)*

## Abstract

Many analysis and prediction tasks require the extraction of structured data from unstructured texts. However, an annotation scheme and a training dataset have not been available for training machine learning models to mine structured data from text without special templates and patterns. To solve it, this paper presents an end-to-end machine learning pipeline, Text2Struct, including a text annotation scheme, training data processing, and machine learning implementation. We formulated the mining problem as the extraction of metrics and units associated with numerals in the text. Text2Struct was trained and evaluated using an annotated text dataset collected from abstracts of medical publications regarding thrombectomy. In terms of prediction performance, a dice coefficient of 0.82 was achieved on the test dataset. By random sampling, most predicted relations between numerals and entities were well matched to the ground-truth annotations. These results show that Text2Struct is viable for the mining of structured data from text without special templates or patterns. It is anticipated to further improve the pipeline by expanding the dataset and investigating other machine learning models. A code demonstration can be found at: https://github.com/zcc861007/Text2Struct

[*]*Correspondence should be addressed*: chaochao.zhou@northwestern.edu

## 1. Introduction

Mining of structured data from text is a fundamental natural language processing task for further quantitative analysis or prediction, whereas nowadays it still heavily relies on manual extraction. For example, meta-analysis is a widely used statistical analysis combining results from multiple published scientific studies [1]. It can compensate for insufficient statistical power due to the small size of samples in a single study. Furthermore, in clinical studies, the evolution of comorbidity and risk factors of chronic illness or disability need to be analyzed based on a big database of patients' diagnosis records [2]. However, both meta-analysis and tracking of illness history require researchers to extract and collect a large amount of data by reviewing numerous texts. Such scanning processes are tedious and time-consuming, so it is desirable to develop an automated pipeline to do the task.

The structured data can be thought of as tabular data with fields and values in relational databases or spreadsheets. For clinical and healthcare studies, quantitative data about outcome measures of patients are preferentially extracted for meta-analysis or other statistical analysis. To that end, mining numeral-relevant structured data was a focus of this work. Therefore, the problem of mining structured data could be formulated as the extraction of metrics and units related to each numeral in the text. For general text without special templates or patterns, there are several challenges in solving this problem. First, multiple numerals may co-exist in a single sentence, so the relations of entities corresponding to each numeral need to be identified. Second, the lengths of words to describe entities such as metrics and units vary in different sentences. In addition, a numeral may be associated with multiple metrics hierarchically.

The overarching goal of this study is to develop an end-to-end machine learning pipeline to automatically mine metrics and units associated with each numeral in text. Concretely, the main works include: 1) collecting texts and performing text pre-processing; 2) creating a dataset with text annotation of entities and relations; 3) developing and training a recurrent neural network (RNN); 4) testing RNN performance.

## 2. Related Work

The semantic relationships between entities in a text can be mined in a manner of joint extraction of entities and their relations [3]. Appropriate tagging plays a key role in the extraction of entities and relations. For example, as shown in **Fig. 1**, two relations such as "Country-President" (CP) and "Company-Founder" (CF) exist in the sentence. Zheng et al. [3] proposed a tagging scheme, in which the relations such as CP and CF were included in the tags of corresponding words, while other irrelevant words were tagged as "O" (other words). Based on the tagging scheme, the output tags with specific meanings can be exactly assigned to each word in the input sentence. Such tagging enables the training of a variety of neural network models in an end-to-end manner. For example, BERT is a transformer model that was pre-trained in two unsupervised text mining tasks using a large-scale general text corpus from English Wikipedia (2,500M words), so it could encode enriched semantic relationships [4]. Recently, pre-trained BERT has been fine-tuned for jointly extracting entities and relations based on the tagging scheme [5]. In particular, a softmax layer was added to BERT as the output layer to predict the tags.

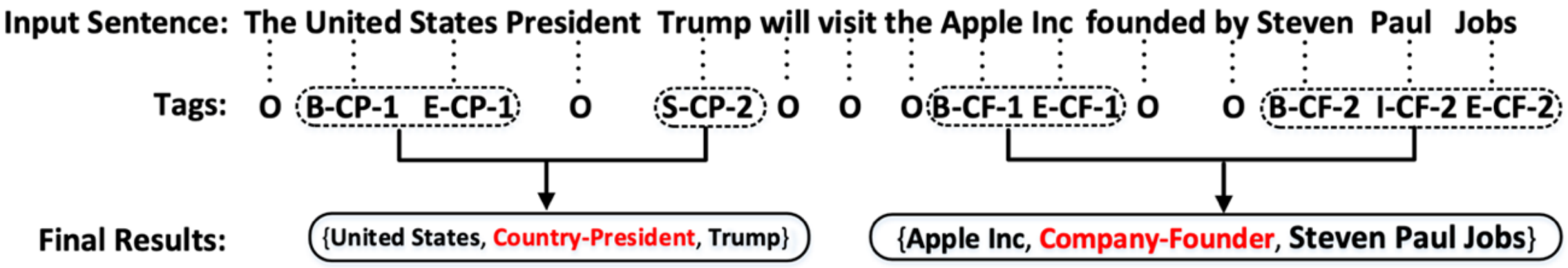

***Figure 1**. A tagging scheme proposed by Zheng et al.* [3]*, where "CP" is short for "Country-President" and "CF" is short for "Company-Founder". "1" and "2" denotes two entities in a relationship. If an entity only includes a single word, it is denoted by "S"; otherwise, it is marked by "B" = Begin, "I" = intermediate, and "E" = End. Additionally, "O" represents other words. (This figure was adopted from* [3]*)*

In another application, BERT was fine-tuned to predict if there is a reasonable semantic relationship between two entities in a sentence [6]. In this work, a sentence template describing a relation between two entities was given, as represented by $\phi(X, Y)$. For example, a template $\phi(X, Y)$ could be "X is the capital of Y". By assigning "Rome" and "Italy" to X and Y in the template, respectively, we can obtain an instance, i.e., "Rome is the capital of Italy", which can be considered as a positive relation. In contrast, "Rome is the capital of France" and "Trump is the capital of Obama" are not true, so both of them should be accounted for as negative relations. Therefore, the problem can be formulated as a binary classification. Given different sentence templates and different pairs of entities, a BERT model can be implemented to predict if there are positive and negative relations. Correspondingly, the output layer of the pre-trained BERT model was modified to a classification layer with linear activation and trained using the binary-cross entropy loss [6].

Furthermore, mining of the relationship between numeral and target entities has also gained increasing attention [7]. In this type of mining, studies have attempted to predict the probability distribution of a value, given a sentence where the numeral was masked [8]. The numeric values can be estimated by learning the distribution of the magnitude of a target value, using either regression or classification (e.g., the values can be mapped to uniformly spaced buckets). For example, given the sentence "The size of the dog is 80 cm.", learning can be performed to predict how large a dog is, using a template "the size of the A is X cm". Likewise, the pre-trained BERT can be adapted to tackle the problem. However, for model training, numerals need to be tokenized to facilitate parsing the natural representation as Arabic numbers. In this work [8], every number in training instances was represented by a scientific notation, i.e., a combination of a 10-based exponent and mantissa. For example, 314.1 ($= 3.141 \times 10^2$) is represented as 3141[EXP]2, where [EXP] is a new token introduced into the vocabulary.

In addition, Chen et al. [9] tackled the numeral attachment problem by developing a dataset and a machine learning model. In their work, they aimed to extract the association of numerals with companies' stocks (denoted by cashtags) from financial tweets. For example, given a tweet "Guess who sold off about $800 million in $MDLZ after losing about $1 billion on $VRX?", "800 million" and "1 billion" are related to $MDLZ and $VRX, respectively, where $MDLZ and $VRX are cashtags frequently used in financial tweets. Given one instance with the target numeral and cashtag (i.e., they used tweet word tokens and a numeral-cashtag pair as model inputs), they formulated the problem as a binary classification problem to determine if the given numeral is related to the given cashtag. Furthermore, they introduced two auxiliary tasks including reason detection and fine-grained reason type classification to improve the performance of numeral attachment predictions.

## 3. Methods

It is expected that an annotated text dataset can be leveraged by contemporary machine learning models to mine structured data from general texts without special templates or patterns. Therefore, we developed the pipeline, Text2Struct, comprising a text annotation scheme (*Section 3.1*), training data processing (*Section 3.2*), and implementation of different machine learning models (*Section 3.3*).

### 3.1. Annotation of Entities and Relations

We annotated the relations between numerals and entities in each sentence using the BRAT annotation tool [10]. In BRAT, the character positions of marked entities are recorded. To ensure that character positions are conveniently and accurately converted to word positions (as we used unigram word-based learning in this work), word segmentation has been used to pre-process each sentence prior to annotation using BRAT. Meanwhile, all numerals including percentages have been converted to floats in the pre-processing (e.g., "50%" is converted to "0.5").

As shown in **Fig. 2**, given a sentence, we aim to annotate the metric and unit associated with each numeral. Some rules have been applied in our annotation. In terms of the numeral-unit relations, it is straightforward to identify the unit of a numeral, but we additionally considered the objects of percentages as a special case of units. For example, in the sentence of **Fig. 2**, we considered "patients" as the unit of these percentages.

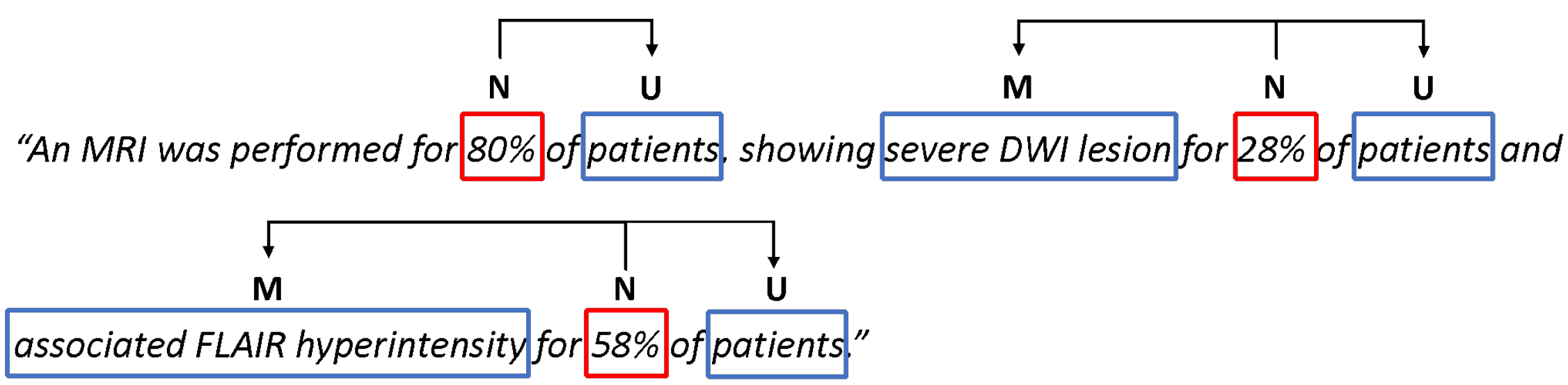

***Figure 2**. An example to annotate entities including metrics (M) and units (U) associated with numerals (N).*

For the numeral-metric relations, there may be multiple metrics corresponding to a single numeral in hierarchical relations. In this study, we focused on the mining of the closest / most direct metric associated with a numeral. Particularly, we prioritized annotating the metric of a numeral, if it is located in the same clause where the numeral lies. For example, in **Fig. 2**, "28%" and "58%" correspond to the metrics of "severe DWI lesion" and "associated FLAIR hyperintensity", respectively. However, it can be further argued that both “28%” and “58%” belong to “80% of patients”. For higher relations like this, we have omitted them in our annotation. Moreover, it should be noted that the numeral-metric relation is not limited to equality. For example, inequalities (e.g., “$p < 0.05$”) and ranges (e.g., “95% CI: 0.31 to 9.53”) are also considered in the numeral-metric relations.

### 3.2. Training Data

The texts and annotated entities in both numeral-unit and numeral-metric relations should be converted to input (text) - output (label) instances to enable subsequent machine learning. As model inputs, texts were represented by index sequence; the index of a word is the ranking of term frequency of the word in the vocabulary built from all annotated texts. Because a sentence often contains multiple numerals, a machine learning model needs to identify the associated entities of each numeral one by one. Therefore, we split such a sentence into multiple training instances, each with one target numeral, while other numerals that are not focused on were masked by “[num]”, as shown in row 1 of **Fig. 3**. Furthermore, the target number was converted to unigram characters while introducing “[neg]” and “[dot]” (because the dash and dot signs have other meanings), as shown in row 2 of **Fig. 3**. The conversion of the target number is based on two considerations. First, different numerals inevitably expand the vocabulary size, but the character representation of numerals only introduces twelve new word tokens, i.e., {0 ~ 9}, [neg], and [dot] (recall that all numerals have been converted to floats in *Section 3.1*). More importantly, the character representation can suggest the magnitude of a numeral, because it can be interpreted by the RNN according to the order of the numeral characters. In contrast, the word representation of numerals is less informative to the RNN.

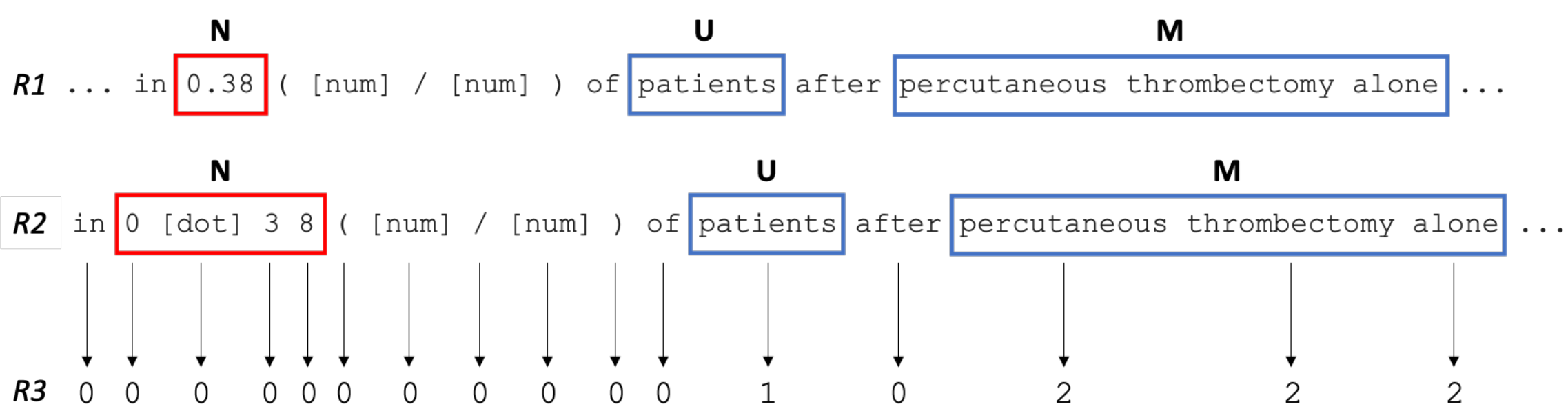

***Figure 3****. A text chunk and its corresponding label converted from annotation. Note that the target numeral in the original text (row 1) has been converted to unigram characters (row 2). The resulting label of the text chunk is presented in row 3 (0 = none, 1 = unit, and 2 = metric).*

The labels of each sentence were created by denoting the tagged units and metrics associated with the target numeral as integers (e.g., we defined that 0 = none, 1 = unit, and 2 = metric), as shown in row 3 of **Fig. 3**. The labels are considered as the ground-truths of the entities and relations, which we expect the model predictions to match. It should be noted that it is not necessary to label the target numeral, as it has been indicated in the input using the character representation (while other irrelevant numerals were denoted by "[num]").

### 3.3. Model Implementation

In our text corpus, some sentences are very long (more than 50 words), while the "effective" ranges of a training instance (i.e., the minimal range where a target numeral and its associated unit and metric occur simultaneously) in these sentences are relatively narrow. Therefore, we shortened the sentence in each training instance by truncating the text portion 5 words before and after the "effective" range. The maximum length of the resulting instances is 48 words after the target numerals were split into characters. Subsequently, all training instances were padded to 50 words.

To mine the units and metrics associated with numerals in text, we adopted an RNN with a many-to-many architecture that was used for English-to-French translation (https://github.com/tommytracey/AIND-Capstone). The index sequences of padded texts with a dimension of 50 were fed to the RNN, sequentially consisting of an embedding layer (with an embedding vector dimension of 128), two bidirectional gated recurrent unit (GRU) layers, a dropout layer, and a dense layer with softmax activation, resulting in an output array with a dimension of (50, 3), equal to the dimension of the one-hot encoding of a ground-truth label. To train the RNN, the sparse categorical cross-entropy was used as the loss function.

As the unit and metric could be sparsely distributed in a label, it is known that the traditional accuracy metric may not be an appropriate evaluation measure for such unbalanced training instances. Therefore, to evaluate the performance of the RNN, we introduced the multi-class soft dice similarity coefficient, which is widely used in the evaluation of image segmentation. Here, the dice coefficient is used to quantify the overlap between the probability prediction ($p$), which is the output of the dense layer with softmax activation, and the one-hot encoding of the ground-truth label ($t$):

$$Dice(p,t) = \frac{1}{N}\sum_{c=1}^{N}\frac{2 \times \sum_i \ p_{i,c}t_{i,c} + \epsilon}{\sum_i \ p_{i,c}^2 + \sum_i \ t_{i,c}^2 + \epsilon} \tag{1}$$

where both $p$ and $t$ have a dimension of (50, 3). $i$ is the index of the prediction/label sequence, and $c$ is the index of classes (in total, there are $N$ = 3 classes, corresponding to three label values, 0 = none, 1 = unit, and 2 = metric). $\epsilon = 1 \times 10^{-5}$ is added to avoid an undefined fraction. The dice coefficient has a range between 0 and 1, and a larger value indicates better performance.

## 4. Experiment

### 4.1. Text Collection

In this work, we demonstrate that the proposed Text2Struct pipeline is able to automatically mine the relationships between numerals and entities (units and metrics) from medical publications. First, we collected abstracts of publications in the recent year (from Nov 2021 to Nov 2022) by searching the PubMed database (https://pubmed.ncbi.nlm.nih.gov/) using a single keyword of "thrombectomy". From the obtained 2451 abstracts, we chose 100 abstracts of the latest publications, and segmented each abstract into sentences. Then, we filtered sentences that include numerals, resulting in 521 sentences. Further, pre-processing such as word segmentation and conversion of numerals to floats as described in *Section 3.1* were conducted for these sentences prior to annotation.

### 4.2. Text Annotation

Based on the annotation rules described in *Section 3.1*, we used the BRAT annotation tool to annotate relations of numerals with units and metrics in the 521 sentences with numerals. Some representative annotation examples are presented in **Fig. 4**. For the first sentence, it is easy to find the corresponding units and metrics of numerals, as they are close to numerals. The second sentence includes a metric ("Five non-inferiority margins") that is far away from numerals. The third sentence is an example where there are hierarchical relations of a numeral with different metrics (e.g., "0.837" corresponds to "large artery atherosclerosis" and further belongs to "major etiologic risk factors"). As discussed above in *Section 3.1*, we only marked the closest metric for each numeral according to our rule. The fourth sentence is an example where both the unit and metric of a numeric exist; that is, the unit and metric of "817" are "participants" and "EVT alone", respectively. In contrast, the last sentence is an example where the numerals, such as years, have no associated entities.

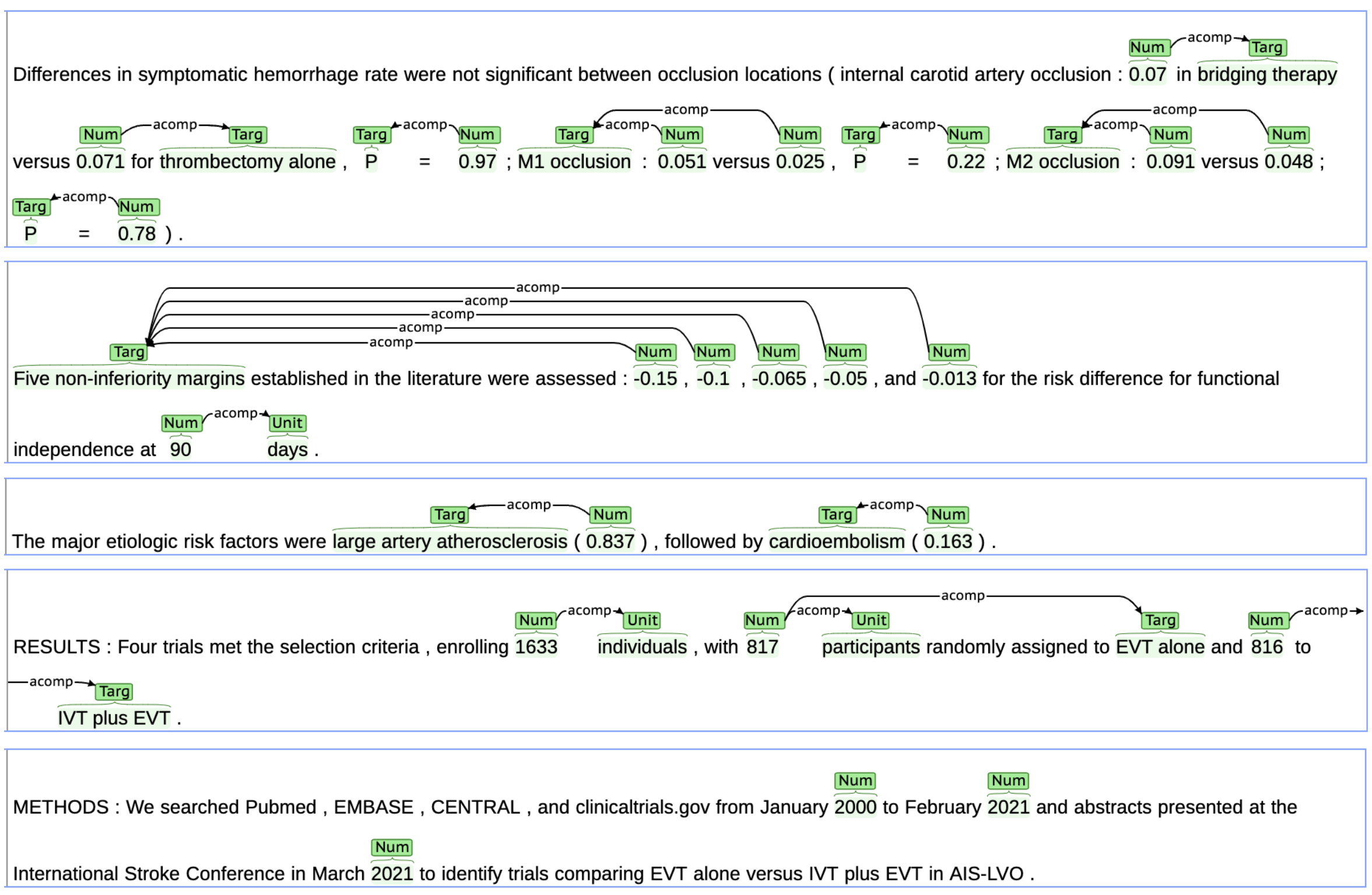

***Figure 4***. *Five selected annotation examples. The tags "Num", "Unit", and "Targ" are used to denote the entities of numerals, units, and metrics, respectively. The association of an entity with a numeral is marked by an arrow pointing from the numeral to the entity.*

### 4.3. Model Training and Testing

After annotation, the annotated 521 sentences were expanded to 1758 individual training instances (i.e., text-label pairs) with single target numerals. Using a ratio of 9:1, these instances were randomly split into a training dataset with 1582 instances and a testing dataset with 176 instances. We trained the RNN on the training dataset using the Adam optimizer with a learning rate of 0.003 and a batch size of 32. In total, 20 epochs were performed according to cross-validation, in which the training loss was below the validation loss at epoch 8.

The model performance to mine associated entities with numerals was tested on the test dataset. The dice coefficient (**Eq. 1**) on the test dataset was 0.82 (by comparison, the accuracy metric was 0.98, which may not properly assess the model performance due to the sparse values of the unit and metric in labels). Some randomly sampled predictions from the test dataset are demonstrated in **Fig. 5**. In the first example, both the unit and metric of the target numeral are well predicted, except that the metric includes an additional word ("was").

In the second example, the unit is not predicted, and the predicted metric has a greater range than the ground truth. In the third example, for the ground truth, there is no unit, and the metric is "[num] CI" (meaning 95% confidence interval), while the prediction is relatively accurate. In the last example, both the unit and metric were not annotated in the ground truth, with which the prediction is exactly coincident.

```
Text #145 is: results : the mean age was 67.6 +/- [num] years ( range , [num]
-
Num is: 67.6
---Ground-Truth---
Unit is: years
Targ is: mean age
---Prediction---
Unit is: years
Targ is: mean age was
```

```
Text #58 is: technical success was achieved in 0.38 ( [num] / [num] ) of
patients after percutaneous thrombectomy alone and in [num] after additional
Num is: 0.38
---Ground-Truth---
Unit is: patients
Targ is: percutaneous thrombectomy alone
---Prediction---
Unit is:
Targ is: ) of patients after percutaneous thrombectomy alone
```

```
Text #144 is: odds ratios : [num] ( [num] CI , 0.62 - [num] ) for internal
Num is: 0.62
---Ground-Truth---
Unit is:
Targ is: [num] CI
---Prediction---
Unit is:
Targ is: ( [num] CI
```

```
Text #91 is: - [num] : [num] versus 0.61 ; p < [num] )
Num is: 0.61
---Ground-Truth---
Unit is:
Targ is:
---Prediction---
Unit is:
Targ is:
```

***Figure 5***. *Randomly sampled predictions compared to the ground truths in the test dataset. In each example, "Num" represents the target numeral, while "Unit" and "Targ" in the ground truth and prediction represent the unit and metric corresponding to the target numeral, respectively.*

## 5. Summary

We demonstrated the feasibility of Text2Struct for mining structured data from text, as supported by the encouraging experimental results. Although we collected abstracts of medical publications regarding thrombectomy as our dataset for model training and testing, they are general texts without the special templates or patterns as defined in most previous works (see our review in *Section 2*). Furthermore, our annotation scheme is very simple with only three tags to mark numerals, units, and metrics, as well as one-way arrows to mark both the numeral-unit and numeral-metric relations. Therefore, it can ease the workload to further expand the annotated text dataset. In particular, we introduced a strategy to extract the closest metric associated with a numeral, whereby hierarchical numeral-metric relations can be mined recursively. That is, once an inner relation is mined, outer relations can be further mined by masking the metric in the inner relation. In addition, Text2Struct can be combined with text retrieval implementations (e.g., BM25 [11]) to filter and collect clinically significant measures, such as NIH Stroke Scale (NIHSS) and Modified Rankin Scale (MRS), which are widely used to assess thrombectomy outcomes. These measures can be mined by Text2Struct with high accuracy.

## 6. Future Work

Further work can be done to improve Text2Struct in several aspects. First, for this study, a small text dataset was created from the literature about "thrombectomy". The ground truths were built by the authors, so there could be uncertainty and inconsistency due to the ambiguity in the writing of the original text, complex hierarchical relations, or limited knowledge about measures in thrombectomy. Hence, it is necessary to expand the dataset and consider other domains, while collecting high-quality and legible texts. Currently, the sentence in each training instance is shortened to the text portion 5 words before and after the "effective" range where the target numeral, unit, and metric occur. Consequently, it limited the length of the input texts fed to the RNN. We will perform statistics regarding the farthest positions of associated entities before and after the target numerals to determine a more appropriate truncated range. Moreover, this work only implemented an RNN for joint relation extraction of numerals and associated entities. To achieve better performance, we will investigate other machine learning models, such as BERT, which encoded rich semantic information by pre-training on a large text corpus.

## Conflicts of Interest

Both authors declare that there is no conflict of interest.

## Reference

[1] M. J. Page *et al.*, "The PRISMA 2020 statement: an updated guideline for reporting systematic reviews," *Syst Rev*, vol. 10, no. 1, Dec. 2021, doi: 10.1186/s13643-021-01626-4.

[2] J. B. Starren, A. Q. Winter, and D. M. Lloyd-Jones, "Enabling a Learning Health System through a Unified Enterprise Data Warehouse: The Experience of the Northwestern University Clinical and Translational Sciences (NUCATS) Institute," *Clin Transl Sci*, vol. 8, no. 4, pp. 269–271, Aug. 2015, doi: 10.1111/cts.12294.

[3] S. Zheng, F. Wang, H. Bao, Y. Hao, P. Zhou, and B. Xu, "Joint Extraction of Entities and Relations Based on a Novel Tagging Scheme," Jun. 2017, [Online]. Available: http://arxiv.org/abs/1706.05075

[4] J. Devlin, M.-W. Chang, K. Lee, and K. Toutanova, "BERT: Pre-training of Deep Bidirectional Transformers for Language Understanding," Oct. 2018, [Online]. Available: http://arxiv.org/abs/1810.04805

[5] B. Qiao, Z. Zou, Y. Huang, K. Fang, X. Zhu, and Y. Chen, "A joint model for entity and relation extraction based on BERT," *Neural Comput Appl*, vol. 34, no. 5, pp. 3471–3481, Mar. 2022, doi: 10.1007/s00521-021-05815-z.

[6] Z. Bouraoui, J. Camacho-Collados, and S. Schockaert, "Inducing Relational Knowledge from BERT," 2020. [Online]. Available: www.aaai.org

[7] M. Yoshida and K. Kita, "Mining Numbers in Text: A Survey," in *Information Systems - Intelligent Information Processing Systems, Natural Language Processing, Affective Computing and Artificial Intelligence, and an Attempt to Build a Conversational Nursing Robot*, IntechOpen, 2021. doi: 10.5772/intechopen.98540.

[8] X. Zhang, D. Ramachandran, I. Tenney, Y. Elazar, and D. Roth, "Do Language Embeddings Capture Scales?," Oct. 2020, [Online]. Available: http://arxiv.org/abs/2010.05345

[9] C. C. Chen, H. H. Huang, and H. H. Chen, "Numeral attachment with auxiliary tasks," in *SIGIR 2019 - Proceedings of the 42nd International ACM SIGIR Conference on Research and Development in Information Retrieval*, Jul. 2019, pp. 1161–1164. doi: 10.1145/3331184.3331361.

[10] P. Stenetorp, S. Pyysalo, G. Topi, T. Ohta, S. Ananiadou, and ichi Tsujii, “BRAT: a Web-based Tool for NLP-Assisted Text Annotation,” *Proceedings of the Demonstrations at the 13th Conference of the European Chapter of the Association for Computational Linguistics*, 2012, [Online]. Available: http://brat.nlplab.org

[11] S. E. Robertson and S. Walker, “Some Simple Effective Approximations to the 2-Poisson Model for Probabilistic Weighted Retrieval,” 1994.